\documentclass[preprint,aps,prd,10pt,superscriptaddress,nofootinbib,showpacs]{revtex4-1}
\usepackage{hyperref}
\usepackage{graphicx}
\usepackage{amsmath,amssymb}
\usepackage{subfigure,dcolumn}
\usepackage[T2A,T1]{fontenc}
\usepackage[russian,english]{babel}
\usepackage{epstopdf}
\usepackage[dvipsnames]{xcolor}
\usepackage{color}
\usepackage{multirow,booktabs}
\usepackage{lipsum}
\usepackage{lineno}
\usepackage{rotating,url}

\usepackage{listings}
\newcommand\figcaption{\def\@captype{figure}\caption}
\newcommand\tabcaption{\def\@captype{table}\caption}
\makeatother

\newcommand{\bra}[1]{\left\langle #1 \right|}
\newcommand{\ket}[1]{\left|#1\right\rangle}

\newcommand{\Dmq}{\Delta m^2}

\newcommand{\eVq}{\ensuremath{\text{eV}^2}}

\usepackage{xcolor}
\usepackage[normalem]{ulem} 
\newcommand\hll{\bgroup\markoverwith
	{\textcolor{yellow}{\rule[-.5ex]{2pt}{2.5ex}}}\ULon}

\begin{document}

\title{Probing the doubly-charged Higgs with Muonium to Antimuonium Conversion Experiment}

\author{Chengcheng Han}
 \email{hanchch@mail.sysu.edu.cn}
 \affiliation{
School of Physics, Sun Yat-Sen University, Guangzhou 510275, China
}
\author{Da Huang}
 \email{dahuang@bao.ac.cn}
\affiliation{
National Astronomical Observatories, Chinese Academy of Sciences, Beijing, 100012, China
}
\affiliation{School of Fundamental Physics and Mathematical Sciences, Hangzhou Institute for Advanced Study,
University of Chinese Academy of Sciences, Hangzhou 310024, China}
\affiliation{International Center for Theoretical Physics Asia-Pacific, Beijing/Hangzhou, China}
 \author{Jian Tang}%
 \email{tangjian5@mail.sysu.edu.cn}
\affiliation{
School of Physics, Sun Yat-Sen University, Guangzhou 510275, China
}%

 \author{Yu Zhang}
 \email{dayu@ahu.edu.cn}
\affiliation{
Institutes of Physical Science and Information Technology, Anhui University, Hefei 230601, China
}
\affiliation{
School of Physics and Materials Science, Anhui University, Hefei 230601,China
}

\date{\today}

\begin{abstract}
The spontaneous muonium-to-antimuonium conversion is one of the interesting charged lepton flavor violation processes. MACE is the next generation experiment to probe such a phenomenon. In models with a triplet Higgs to generate neutrino masses, such as Type-II seesaw and its variant, this process can be induced by the doubly-charged Higgs contained in it. In this article, we study the prospect of MACE to probe these models via the muonium-to-antimuonium transitions. After considering the limits from $\mu^+ \rightarrow e^+ \gamma $ and $\mu^+ \rightarrow  e^+ e^- e^+$, we find that MACE could probe a parameter space for the doubly-charged Higgs which is beyond the reach of LHC and other flavor experiments. 

\end{abstract}

\maketitle
\section{Introduction}

The observation of neutrino oscillations has indicated that neutrinos have very tiny but non-zero masses, which is one of the direct evidences towards the new physics (NP) beyond the Standard Model (SM). Traditionally, the explanation of the measured neutrino masses and mixings involves the seesaw mechanism where the smallness of neutrino masses is allowed with the existence of heavy eigenstates. One of the most striking predictions of the seesaw models is the existence of charged lepton flavor violation (cLFV) processes. Indeed, the neutrino oscillations can be viewed as a manifestation of the neutral lepton flavor violation (LFV). Therefore, it is natural to think about the corresponding LFV in the charged lepton channels. Currently, there are many ongoing and up-coming experiments searching for the cLFV processes all over the world. The accelerator muon beam experiments at PSI are searching for $\mu^+\to e^+ e^+ e^-$ by the Mu3e Collaboration~\cite{Berger:2014vba} and for $\mu^+ \to e^+ \gamma$ by MEG-II~\cite{Baldini:2018nnn}. Furthermore,  the coherent muon-to-electron conversions ($\mu^-N\to e^- N$) will be searched for by COMET~\cite{Adamov:2018vin} at J-PARC and Mu2e~\cite{Bartoszek:2014mya} at FNAL, both of which are still under construction. 

Another equally important but less studied cLFV channel is the muonium-to-antimuonium conversion, which was first brought forward by Pontecorvo~\cite{Pontecorvo:1957cp} more than half a century ago. Muonium $M$ is a hydrogen-like bound state ($e^-\mu^+$) formed by an electron and an antimuon, while an antimuonium $\bar{M}$ is the corresponding anti-particle bounded by a positron and a muon ($e^+\mu^-$). Experimentally, one can produce muonium atoms by colliding slow muons into a target material, and then tries to see the spontaneous conversion from muoniums into antimuoniums. In the SM, such a conversion is forbidden by the lepton flavor symmetry. Hence, the detection of this process can be viewed as a probe to the physics beyond the SM. In light of its potential significance, there have been many theoretical studies on the muonium-to-antimuonium transition in some well-motivated models, such as the type-I~\cite{Clark:2003tv,Cvetic:2005gx,Liu:2008it} and type-II seesaw~\cite{Chang:1989uk} models, the $Z_8$ model with more than 4 lepton generations~\cite{Fujii:1993su}, the minimal
$331$ model~\cite{Frampton:1997in}, the R-parity violated
SUSY model~\cite{Halprin:1993zv} and left-right symmetric model~\cite{Herczeg:1992pt}, as well as the low-energy effective operator framework~\cite{Halprin:1982wm}. More recently, the $M-\bar{M}$ conversion was explored theoretically in a model-independent way in Ref.~\cite{Conlin:2020veq}.

At present, the best upper constraint on the muonium-to-antimuonium conversion has been given by the PSI experiment in 1999, which presented the upper limit on the conversion probability to be ${\cal P}(M\leftrightarrow \bar{M})\lesssim 8.3\times10^{-11}$ at the 90\% confidence level~\cite{Willmann:1998gd}. During the past two decades, there has not been any experimental improvement in this important cLFV channel. However, this situation is expected to be soon changed in the near future due to the advent of the Muonium-to-Antimuonium Conversion Experiment (MACE) in China~\cite{MACE:2020}. MACE is the next-generation experiment specialized to measure this exceptional cLFV process, which will be operated at the upgraded 500~kW pulsed proton accelerator running at the China Spallation Neutron Source (CSNS). By taking advantage of the high-quality intense slow muon sources with more than $10^8$ $\mu^+$ produced per second and the beam spread smaller than 5\% at CSNS, together with the high-efficiency muonium formation targets, the high-precision magnetic spectrometer, and the optimized detector setup by locating the detector at the front end of a future muon collider, MACE is expected to enhance the sensitivity to the $M-\bar{M}$ conversion probability by more than two orders of magnitude from the existing PSI experiment. Therefore, we expect that MACE can play an important role in probing and constraining the parameter space of NP models to generate the neutrino masses.

In light of promising experimental developments projected for by MACE, it is timely to investigate the muonium-to-antimuonium conversion in more detail, paying attention to its interplay with other flavor and collider searches in probing the model parameter space of interest. In the present paper, we shall explore this aspect in the type-II seesaw model and its variant as a case study. The traditional type-II seesaw model generates the small active neutrino masses by introducing a $SU(2)_L$ triplet scalar, in which the doubly-charged scalar component can induce the muonium-to-antimuonium conversion at tree level. Note that, due to the model structure in the type-II seesaw, the predicted $M-\bar{M}$ conversion probability is intimately correlated to the neutrino oscillation data and other cLFV processes, such as $\mu^+ \to e^- e^+ e^+$ and $\mu^+ \to e^+\gamma$. As a result, it will be seen below that MACE is not sensitive to the relevant parameter space surviving after imposing the existing cFLV constraints. In other words, if MACE can observe the positive signal on the muonium-to-antimuonium transition, it means that the simplest type-II seesaw model is ruled out and one is required to consider some extensions to it. In our discussion, we shall provide such an example by incorporating in the type-II seesaw model only a single heavy right-handed neutrino, which will be called the hybrid seesaw model in the following. It turns out that MACE can provide additional information in this hybrid model on the parameter regions which cannot be reached by other experiments.  

This article is organized as follows: in Sec.~\ref{sec:QM}, we briefly describe quantum mechanics of the muonium-to-antimuonium transition based on the effective field theory, and then estimate the MACE sensitivity to this process. In Sec.~\ref{sec:seesaw} we give a brief introduction to the type-II and hybrid seesaw models. Then we detail the leading-order calculation of the $M-\bar{M}$ conversion probability in these two models. Sec.~\ref{sec:flavor} is devoted to the discussion of existing constraints on the type-II and hybrid seesaw models, including those from the measured neutrino masses and mixings, other cLFV processes, and collider searches on the doubly-charged scalar. We present our numerical results in Sec.~\ref{sec:calc}. Finally, we summarize and conclude in Sec.~\ref{sec:conclusion}.

\section{MACE prospects on the muonium-antimuonium transitions}
\label{sec:QM} 

Muonium $\ket{M}$ is a non-relativistic Coulombic bound state of
$\mu^+$ and $e^-$, and antimuonium $\ket{\overline{M}}$ is a similar
bound state of $\mu^-$ and $e^+$. The nontrivial mixing between
$\ket{M}$ and $\ket{\overline{M}}$ implies the non-vanishing
Lepton Flavor Violation(LFV) amplitude for $e^-\mu^+\to e^+\mu^-$.

Currently, the most precise measurement of the conversion probability of the $M-\bar{M}$ conversion was given by the PSI Collaboration two decades ago, with $P (M\leftrightarrow \overline{M}) < 8.3\times 10^{-11}$ at 95$\%$ C.L.. The MACE Collaboration at CSNS attempts to improve the sensitivity of this conversion to the level of $P(M\leftrightarrow \overline{M}) \sim {\cal O}(10^{-14})$. 

Microscopically, the transition between $M$ and $\overline{M}$ can be described by the following local effective Hamiltonion density~\cite{Cvetic:2005gx}:
\begin{equation}\label{EffHam}
  \mathcal{H}_{\mbox{eff}}=\frac{G_{M\overline{M}}}{\sqrt{2}}[\bar{\mu}(x)\gamma^\alpha(1-\gamma^5)e(x)]
  [\bar{\mu}(x)\gamma_\alpha(1-\gamma^5)e(x)]\,,
\end{equation}
where $G_{M\overline{M}}$ is the corresponding Wilson coefficient.
Due to the LFV transition, $\ket{M}$ and $\ket{\overline{M}}$ are not
mass eigenstates. In this basis, the mass matrix has non-diagonal
components:
\begin{equation}
  m_{M\overline{M}}=\bra{M}\int d^3x
  \mathcal{H}_{\mbox{eff}}(\vec{x})\ket{\overline{M}}
\end{equation}
which can be transformed into the form of its constituents with a
momentum distribution $f(p)$:
\begin{equation}
  \ket{M(0)}=\int
  \frac{d^3p}{(2\pi)^3}f(p)a_{e}^\dagger(p,s)b_{\mu}^\dagger(-p,s)\ket{0}
\end{equation}
where $a_{e}^\dagger(p,s)$ creates a fermion $e^-$ with energy
$+E_p$ and momentum $\vec{p}$ and $b_{\mu}^\dagger(-p,s)$ creates an
antiparticle $\mu^+$ with the opposite momentum and the same energy
$+E_p$. If we neglect the momentum dependence in the spinors, the mass mixing
element is:
\begin{equation}
  m_{M\overline{M}}=16\times
  \frac{G_{M\overline{M}}}{\sqrt{2}}\left|\int
  \frac{d^3p}{(2\pi)^3}f(p)\right|^2[\bar{u}_L^{(\mu)}\gamma^\alpha u_L^{(e)}\bar{v}_L^{(\mu)}\gamma_\alpha
  v_L^{(e)}]\,.
\end{equation}
Here the integral of $f(p)$ in the module $|\cdots|^2$ is the
spatial wavefunction at zero distance so that
$|\cdots|^2=|\Psi(0)|^2/(2m_em_\mu)$. The latter spinor product can
be simplified into $2m_em_\mu$ in terms of their normalizations and
spin combinations. Therefore, the mass mixing element arrives at:
\begin{equation}
  m_{M\overline{M}}=16\times
  \frac{G_{M\overline{M}}}{\sqrt{2}}\frac{|\Psi(0)|^2}{2}=16\times
  \frac{G_{M\overline{M}}}{\sqrt{2}}\frac{(\mu\alpha)^3}{2\pi}
\end{equation}
with the reduced mass $\mu=\frac{m_em_\mu}{m_e+m_\mu}\approx m_e$.
The mass eigenstates are then the simple combinations $|M_1\rangle =
(|M\rangle + |\bar M\rangle)/\sqrt{2}$ and $|M_2\rangle = (|M\rangle
- |\bar M\rangle)/\sqrt{2}$, and the difference in mass eigenvalues is given by
$\Delta m\equiv$ $m_1 - m_2 = 2\ m_{M\bar M}$. In experiments, muonium oscillation time is much longer 
than their decay time. So one can
only hope to observe the mixing phenomenon by measuring the
probability that a state that starts as a muonium($\mu^+ e^-$)
decays as an antimuonium, where final states include a high energy electron and a low
energy positron. Therefore, the state that starts as $|M\rangle$ at $t=0$
(denoted it by $|M(t)\rangle$) evolves as follows:
\begin{equation}
|M(t)\rangle = |M_1\rangle \langle M_1|M\rangle e^{-i\ m_1 t} +
|M_2\rangle \langle M_2|M\rangle e^{-i\ m_2 t}
\end{equation}
and will decay as a $\bar M$ with a total probability:
\begin{equation}
P(M\to \bar M) = \int_0 ^\infty \frac{dt}{\tau} e^{-t/\tau} |\langle
\bar M|M(t)\rangle|^2 \quad = \frac{(\Delta m\ \tau)^2}{2(1+(\Delta
m\ \tau)^2)} \approx \frac{1}{2} (\Delta m\ \tau)^2
\label{prob-delta}
\end{equation}
where $\tau$ is the muon lifetime. Eq.\ (\ref{prob-delta}) does not
take into account the effect of static electromagnetic fields in
materials, which break the degeneracy $m_{MM} = m_{\bar M\bar M}$
and further suppresses the probability~\cite{Feinberg:1961zza}, an
important effect in some experiments. In any case, we see from Eq.\
(\ref{prob-delta}) that the conversion probability is in general
very small and proportional to $|G_{M\bar M}|^2$:
\begin{equation}
P(M\to \bar M) = \frac{64\ G_F^2\ \alpha^6 m_e^6\ \tau^2}{\pi^2}
\left(\frac{G_{M\bar M}}{G_F}\right)^2\quad =2.64\times 10^{-5}
\left(\frac{G_{M\bar M}}{G_F}\right)^2. \label{probability}
\end{equation}
{Based on this formula for the transition probability, it is easy to transform the PSI limit to that on the Wilson coefficient as $G_{M\overline{M}}/G_F < 3\times 10^{-3}$, while the MACE experiment is expected to improve the sensitivity by at least two orders with $G_{M\overline{M}}/G_F \lesssim {\cal O} (10^{-5})$.}

\section{The Muonium-antimuonium Conversion in Seesaw Models}
\label{sec:seesaw}

Now we explore the charged lepton flavor violation in the Type-II seesaw model with the $M-\overline{M}$ transition. Especially, we would forecast the prospects of the MACE experiment to detect such a muonium-to-antimuonium conversion and constrain the parameter space in this popular neutrino mass model. 

The type-II seesaw model, which is an extension of the standard
model with a weak-scale triplet Higgs boson, is capable of
generating small neutrino masses naturally. First of all, we briefly
review the model in which the triplet Higgs possesses a weak scale
mass, concentrating on how small neutrino masses are produced
\cite{Schechter:1980gr,Kakizaki:2003fc}. The triplet is arranged
into an $SU(2)$ scalar multiplet denoted by $\Delta$ with
hypercharge $Y=1$:
\begin{eqnarray}
  \begin{array}{lll}
  \Delta = \left(
    \begin{array}{cc}
      \xi^+/\sqrt{2} & \xi^{++} \\
      \xi^0 & - \xi^+/\sqrt{2}
    \end{array}
    \right).
  \end{array}
\end{eqnarray}
The standard model gauge symmetry allows the Yukawa interaction
between the lepton doublet $l=(\nu_L,e_L)^T$ and the triplet
$\Delta$:
\begin{eqnarray}
  {\cal L}_\mathrm{Yuk} & = &- \frac{1}{2} (y_N)_{ij}
  \bar{l}^c_i \varepsilon \Delta l_j + \mbox{h.c.} \nonumber \\
  & = &- \frac{1}{2} (y_N)_{ij} \left[
    \bar{\nu}^c_i P_L\nu_j \xi^0
    - \frac{1}{\sqrt{2}} (\bar{\nu}^c_i P_Le_j + \bar{e}^c_i P_L\nu_j) \xi^+
    - \bar{e}^c_i P_Le_j \xi^{++}
    \right]  + \mbox{h.c.},
\label{Yuk}
\end{eqnarray}
where $(y_N)_{ij}$ are Yukawa coupling constants and the Latin
indices $i,j$ represent generations and $\varepsilon\equiv
i\sigma^2$. 

It is easy to get $(y_N)_{ij}=(y_N)_{ji}$ and find that two terms of $\xi^+$ are actually
equal.  The final Yukawa terms for
$\Delta$ arrive at:
\begin{eqnarray}
{\cal L}_\mathrm{Yuk}&=&\frac{1}{ \sqrt{2}} \left[\xi ^+
   \overline{\nu_L^c}\Bigl(U_{\nu }^Ty_N\Bigr)e_L
   +\xi ^{-} \overline{e_L}
   \left(y_N^{*}U_{\nu *}\right)\nu _L^c\right]+\frac{1}{2}
   \left[  \xi^{\text{++}} \overline{e_L^c}\Bigl(y_N\Bigr)e_L
   + \xi ^{--} \overline{e_L} \left(y_N^{*}\right)e_L^c\right] \nonumber\\
   &&-\frac{1}{2} \left[\xi ^{\text{0*}}
   \overline{\nu _L^c}\left(y_N^{*}\right)\nu _L+\xi ^0
   \overline{\nu _L}\Bigl(y_N\Bigr)\nu _L^c\right]-\frac{1}{2}\left[ \overline{\nu _L^c}\Bigl(m_\nu\Bigr)\nu _L
   + \overline{\nu_L^c}\left(v_3^*U_\nu y_N^{*}U_\nu^{T}\right)\nu_L\right]
\end{eqnarray}
In addition to nonzero neutrino masses, the theory predicts the
existence of three neutral, one charged , and one doubly charged
physical Higgs scalar particles. Their masses and couplings with
leptons and quarks depend crucially on the mechanism used to break
the global U(1) symmetry associated with the conservation of lepton
number L. In one realization is that the lepton number L is conserved by the Higgs
potential V($\Phi,\Delta$), while the U(1) global symmetry associated
with the conservation of L is broken spontaneously as $\xi^0$
develops a nonzero vacuum expectation value (VEV). At this time a linear
combination of the imaginary parts of the neutral components of the
Higgs doublet field $\Phi$ will act as a physical massless neutral
scalar particle called a Majoron which is almost ruled out in
experiments. So we do not take this case into account.
Interestingly, the other possibility is to assign two units of L to
the Higgs triplet $\Delta$ ($L_\Delta=-2$) so that the Yukawa
coupling terms($\bar{l}^c\varepsilon\Delta l$) would conserve the
lepton number. The form of the U(1) symmetry breaking leading to
$\langle\xi^0\rangle\neq0$ and L nonconservation is determined by
the assumed properties of the Higgs potential V($\Phi,\Delta$) of
the theory. The details of Higgs potential and the gauge sector can be found in the Appendix.

In addition to the triplet Higgs, we also consider the possibility that the neutrino masses get contributions from the other particles. In particular, we introduce a right-handed neutrino $\nu_R$ with additional couplings:
\begin{eqnarray}
\mathcal{L} \supset -{y_\nu}_i \bar L_i \Phi \nu_R + M_R \nu_R^c \nu_R\,. 
\end{eqnarray}
Then part of the neutrino mass matrix could be generated from the type-I seesaw mechanism. We take this model as the  hybrid seesaw model.

While $\xi^0$ develops a vacuum expectation value $v_3$, the total Majorana
neutrino masses are:
\begin{eqnarray}
  (m_N)_{ij} = (y_N)_{ij}v_3\, + {y_\nu}_i {y_\nu}_j \frac{v^2}{2 M_R},
\end{eqnarray}

The diagonalized neutrino masses is obtained by unitary
transformations:
\begin{equation}
  \nu_L\rightarrow U_\nu \nu_L,\quad
U_\nu^Tm_NU_\nu=\mbox{diag}\{m_1,m_2,m_3\}\equiv m_\nu,\quad
m_i\geq0.%
\label{eqn:diagonalization}
\end{equation}
Without loss of generality, we can work in a basis where charged
lepton masses are diagonalized. Then $U_\nu$ is the PMNS mixing
matrix in charge-current interactions.

Given the Yukawa interaction in Eq.~(\ref{Yuk}), it is easy to draw the leading-order contribution to the muonium-antimuonium conversion. As described in~\cite{Chang:1989uk}, if the
usual Yukawa coupling of the lepton is assumed to be diagonal, we can obtain:
\begin{equation}
\mathcal{H}_{\mathrm{eff}}=\frac{G_{M\bar{M}}}{\sqrt{2}}
[\bar{\mu}\gamma^\mu(1-\gamma^5)e][\bar{\mu}\gamma_\mu(1-\gamma^5)e]\,, \text{with } G_{M\bar{M}}=\frac{(y_{N})_{ee} (y_{N}^* )_{\mu\mu}}{16\sqrt{2}m_{++}^2}\,.
\label{eqn:effective-hamiltonian}
\end{equation}
The integrated probability that the muonium
$M(\mu^+e^-$) decay as $\mu^-$ rather than $\mu^+$ is: 
\begin{equation}
\mathcal{P}(M\to\bar{M})=64^3\bigl(\frac{3\pi^2\alpha^3}{G_Fm_\mu^2}
\bigr)^2 \bigl(\frac{m_e}{m_\mu}
\bigr)^6\bigl(\frac{G_{M\bar{M}}}{G_F} \bigr)^2\,.
\end{equation}
According to the latest experimental upper bound from PSI, we can obtain
$\mathcal{P}(M\to\bar{M})\leq 2.0\times10^5G_{M\bar{M}}^2$ at 90\% confidence level.

\begin{table}\centering
  \begin{footnotesize}
    \begin{tabular}{c|c|c}
      \hline\hline
      & Normal Ordering 
      & Inverted Ordering 
      \\[1mm]
    \hline{}
      \rule{0pt}{4mm}\ignorespaces
       $\sin^2\theta_{12}$
     & $0.269 \to 0.343$
    & $0.269 \to 0.343$
      \\[1mm]
       $\sin^2\theta_{23}$
       & $0.407 \to 0.618$
       & $0.411 \to 0.621$
      \\[1mm]
       $\sin^2\theta_{13}$
       & $0.02034 \to 0.02430$
       & $0.02053 \to 0.02436$
      \\[1mm]
       $\dfrac{\Dmq_{21}}{10^{-5}~\eVq}$
       & $6.82 \to 8.04$
       & $6.82 \to 8.04$
      \\[3mm]
       $\dfrac{\Dmq_{3\ell}}{10^{-3}~\eVq}$
       & $+2.431 \to +2.598$
       & $-2.583 \to -2.412$
      \\[2mm]
      \hline\hline
    \end{tabular}
  \end{footnotesize}
  \caption{
  The allowed ranges of the neutrino
mass square differences and mixing angles at 3$\sigma$ confidence level for Normal Ordering  and Inverted Ordering used in
our numerical analysis \cite{Esteban:2020cvm}.  Note that $\Dmq_{3\ell} \equiv \Dmq_{31} > 0$ for Normal Ordering and
    $\Dmq_{3\ell} \equiv \Dmq_{32} < 0$ for Inverted Ordering. }
  \label{tab:bfranges}
\end{table}

\section{Constraints on seesaw models}
\label{sec:flavor}

The seesaw models considered in the present paper are strongly constrained by the lepton flavor violation (LFV) processes involving the charged lepton sector, such as $\ell_a^\pm \to \ell_b^\mp \ell_c^\pm \ell_d^\pm$ and $\ell^\pm_a \to \ell_b^\pm \gamma$. In both Type-II seesaw and hybrid models, the leading-order contribution to $\ell_a^\pm \to \ell_b^\mp \ell_c^\pm \ell_d^\pm$ is given by the tree-level diagram induced by the doubly-charged scalar $\xi^{++}$. With the Yukawa couplings defined in Eq.~(\ref{Yuk}), the corresponding branching ratios are given as follows~\cite{Primulando:2019evb}:
\begin{eqnarray}
{\cal B} (\ell_a^\pm \to \ell_b^\mp \ell_c^\pm \ell_d^\pm) = \frac{1}{8(1+\delta_{cd})} \frac{|(y_N)_{ab} (y_N^\dagger)_{cd}|^2}{G_F^2 m_{++}^4}\,,
\end{eqnarray}
where the Kronecker delta $\delta_{cd}$ accounts for identical leptons in the final states. In particular, the most precise channel in this class is provided by the decay $\mu^+ \to e^+ e^- e^+$ with the branching ratio given by~\cite{Pal:1983bf,Leontaris:1985qc,Swartz:1989qz,Mohapatra:1992uu,Cirigliano:2004mv,Akeroyd:2009nu, Dinh:2012bp, Dev:2018sel,Ferreira:2019qpf}
\begin{eqnarray}
{\cal B} (\mu^+ \to e^+ e^- e^+) = \frac{|(y_N)_{\mu e} (y_N^\dagger)_{ee}|^2}{16 G_F^2 m_{++}^4}\,,
\end{eqnarray}  
which can be compared with the current best upper bound as follows~\cite{Bellgardt:1987du}
\begin{eqnarray}
{\cal B} (\mu^+ \to e^+ e^- e^+) \leqslant 1.0 \times 10^{-12}\,.
\end{eqnarray}

For the Type-II seesaw model, the LFV decay process $\mu \to e \gamma$ is generated at one-loop level with the help of the doubly- and singly-charged scalars. As a result, the partial width for this process is given by~\cite{Primulando:2019evb, Leontaris:1985qc, Mohapatra:1992uu, Cirigliano:2004mv, Akeroyd:2009nu, Dinh:2012bp, Dev:2018sel, Ferreira:2019qpf}
\begin{eqnarray}\label{meg}
{\cal B} (\mu \to e \gamma) \simeq \frac{\alpha}{768\pi} \frac{\left|(y_N^\dagger y_N)_{e\mu}\right|^2}{G_F^2} \left(\frac{1}{m_+^2} + \frac{8}{m_{++}^2}\right)^2\,,
\end{eqnarray} 
where $\alpha$ refers to the electromagnetic fine structure constant, and in our derivation we have ignored the dependence on the internal lepton masses since they are assumed to be much smaller than the $m_+$ and $m_{++}$. On the other hand, for the hybrid model, the heavy singlet right-handed neutrino would mix with the $SU(2)_L$ active neutrinos and give rise to an additional one-loop contribution to $\mu\to e\gamma$~\cite{Dinh:2012bp,Dev:2018sel,Ferreira:2019qpf,Minkowski:1977sc, Lim:1981kv,Langacker:1988up,Marciano:1977wx,Cheng:1980tp, Ilakovac:1994kj,Deppisch:2004fa,He:2002pva,Lavoura:2003xp,Forero:2011pc,Alonso:2012ji}. Even though this new LFV mode would be enhanced by breaking of GIM mechanism in the SM~\cite{Minkowski:1977sc, Lim:1981kv,Langacker:1988up,Marciano:1977wx,Cheng:1980tp,Ilakovac:1994kj,Deppisch:2004fa, He:2002pva,Lavoura:2003xp,Forero:2011pc,Alonso:2012ji}, it can be shown~\cite{Dinh:2012bp,Dev:2018sel,Ferreira:2019qpf,Forero:2011pc,Alonso:2012ji} that the corresponding amplitude is subdominant compared with those mediated by the charged scalars $\xi^{+},\xi^{++}$ and thus can be safely ignored. Therefore, the dominant contribution to $\mu\to e\gamma$ remains to be the same as that of the Type-II seesaw in Eq.~(\ref{meg}). Nowadays, the most stringent constraint on $\mu \to e\gamma$ is given by the MEG collaboration with the upper bound as follows~\cite{TheMEG:2016wtm}
\begin{eqnarray}
{\cal B} (\mu \to e\gamma) < 4.2 \times 10^{-13}\,.
\end{eqnarray}
Later, we will use it to constrain the parameter spaces of both models in our numerical scanning.

The seesaw models considered in the present paper also suffer constraints from the $\mu \to e$ conversion in nuclei, which can arise from both short-range and long-range contributions. However, explicit calculations~\cite{Cirigliano:2004mv,Alonso:2012ji} have shown that the current $\mu \to e$ conversion sensitivity cannot place any useful constraints on the parameter spaces, given already severe limits given by $\mu \to e \gamma$ and $\mu^+ \to e^+ e^- e^+$. Furthermore, the lepton number conservation in both models is broken by two units and the obtained active neutrino masses are Majorana in nature. Thus, these models can be constrained by the lepton-number violating processes like neutrinoless double beta ($0\nu\beta\beta$) decays in nuclei which have not yet been found so far. In the Type-II seesaw model, there are two contributions to this process: one is the ordinary long-range channel by exchanging the light active neutrinos $\nu$, while the other is the short-range mode mediated by the doubly-charged scalar $\xi^{\pm\pm}$. However, these channels are severely suppressed either by the tiny light neutrino masses or by the doubly-charged scalar mass scale~\cite{Chakrabortty:2012mh,Dev:2018sel}, and thus can be neglected. The inclusion of one heavy singlet right-handed neutrino in the hybrid model does not alleviate the problem since the short-range mode induced by it suffers from a significant suppression arising from the extremely small heavy-light neutrino mixing. In summary, both models are insufficient to generate observable $0\nu\beta\beta$ signals, and thus cannot be constrained by the limits from the existing experiments, like KamLAND-Zen~\cite{KamLAND-Zen:2016pfg} and GERDA~\cite{Agostini:2018tnm}.  

The search for doubly charged Higgs boson has been presented at the LHC with the ATLAS detector \cite{Aaboud:2017qph}, in which the analysis focused 
on the decays $\xi^{\pm\pm}\to e^\pm e^\pm$, $\xi^{\pm\pm}\to e^\pm \mu^\pm$ and $\xi^{\pm\pm}\to \mu^\pm \mu^\pm$.
The partial decay width of $\xi^{\pm\pm}$ to leptons is given by 
\begin{eqnarray}
\Gamma(\xi^{\pm\pm}\to \ell^\pm \ell^{\prime\pm})=k \frac{(y_N)_{\ell \ell^\prime}^2}{16\pi} m_{++},
\end{eqnarray}
where $k=2$ for $\ell = \ell^\prime$ and $k=1$ for $\ell \neq \ell^\prime$.
Assuming $\xi^{\pm\pm}$ almost all decay to leptons, i.e.,
$\sum \limits_{\ell=e,\mu,\tau} {\cal B}(\xi^{\pm\pm}\to \ell^\pm \ell^{\prime\pm})\simeq 1 $, we can get the branch ratios
\begin{eqnarray}
{\cal B} (\xi^{\pm\pm}\to \ell^\pm \ell^{\prime\pm})= \frac{k (y_N)_{\ell \ell^\prime}^2}{\sum \limits_{\ell=e,\mu,\tau}k (y_N)_{\ell \ell^\prime}^2 }.
\end{eqnarray}

\section{Numerical Results}
\label{sec:calc}

\begin{figure}[!t]
\centering
\includegraphics[width=0.4\textwidth]{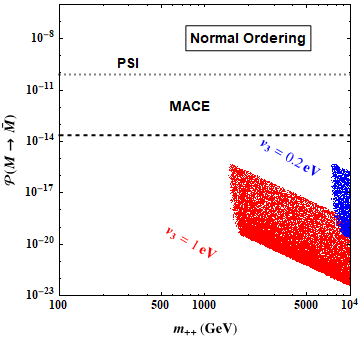}
\includegraphics[width=0.4\textwidth]{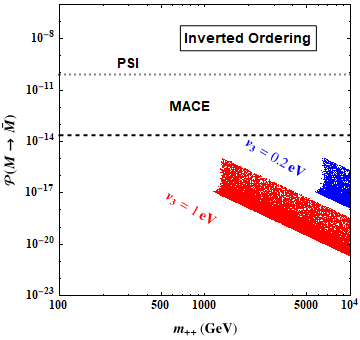}
\includegraphics[width=0.4\textwidth]{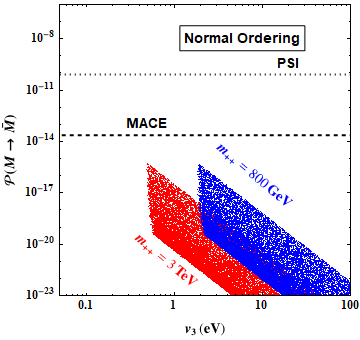}
\includegraphics[width=0.4\textwidth]{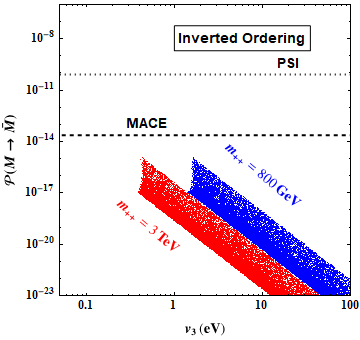}
 \caption{\label{fig:Pwithv1} 
 Upper: Scatter plots of the doubly charged Higgs mass $m_{++}$  and the conversion probability $P$ in the type-II seesaw with $v_3=0.2$ eV (blue points) and  $v_3=1$ eV (red points) for Normal Ordering (left) and Inverted Ordering (right).
 Lower: Scatter plots of the vev $v_3$ and the conversion probability $P$ in the type-II seesaw with $m_{++}=800$ GeV (blue points) and  $m_{++}=3$ TeV (red points) for Normal Ordering (left) and Inverted Ordering (right).
 All the plotted points are allowed by the flavor and collider constraints.
 }
 \label{figure1}
\end{figure}

In this section, we present our numerical results by sampling in the model parameter space. For simplicity, we assume all the model parameters are real.  In our scanning, we selected models that predict the neutrino mass squared differences and mixing angles to be within the experimentally allowed ranges for both normal ordering (NO) and inverted ordering (IO) at $3\sigma$ confidence level obtained from the existing neutrino oscillation data as shown Table~II. Although the recently released data from T2K and NO$\nu$A Collaborations prefer a non-zero Dirac $CP$ phase $\delta_{\rm CP}$, this result is still not conclusive. Thus, we still allow $\delta_{\rm CP}$ to be varied within $[0,\,2\pi]$. We also take into account the cosmological limit on the sum of neutrino masses so that the lightest neutrino mass is sampled uniformly within the range of $[0,\, 0.05\,{\rm eV}]$. The value of the triplet Higgs VEV is adopted to be from $10^{-2}$~eV to $1$~GeV, while the mass doubly-charged Higgs $\xi^{\pm\pm}$ is allowed to be within $[100~{\rm GeV},\, 10~{\rm TeV}]$. Furthermore, we require that the selected models should satisfy the stringent bounds for cLFV processes, such as $\mu\to e\gamma$ and $\mu^+ \to e^- e^+ e^+$. For the remaining model parameters, we compare their predicted muonium-to-antimuonium conversion probability ${\cal P}(M\leftrightarrow \bar{M})$ with the existing PSI bound and the expected MACE sensitivity. In the following, we shall give the final scanning results for the Type-II and hybrid seesaw models, respectively.

\subsection{Type II seesaw}
\begin{figure}[!t]
\centering
\includegraphics[width=0.4\textwidth]{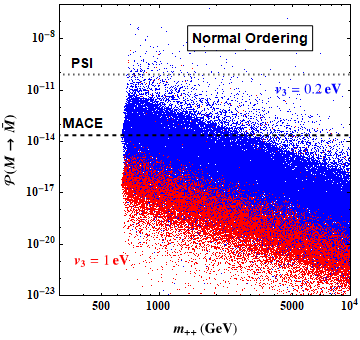}
\includegraphics[width=0.4\textwidth]{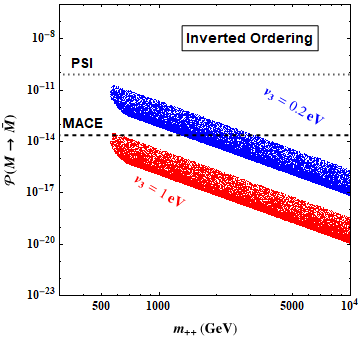}
\includegraphics[width=0.4\textwidth]{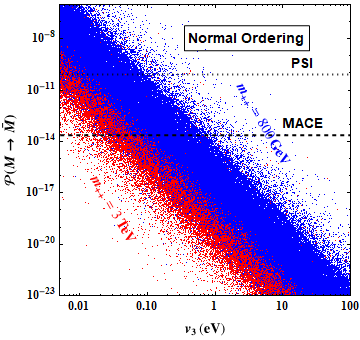}
\includegraphics[width=0.4\textwidth]{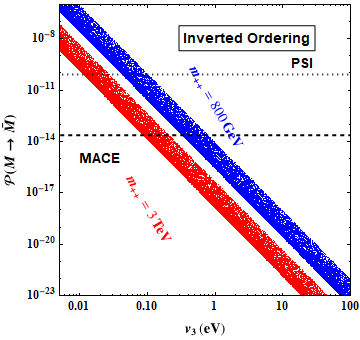}
 \caption{\label{fig:Pwithv2}  Upper: Scatter plots of the doubly charged Higgs mass $m_{++}$  and the conversion probability $P$ in the hybrid seesaw model with $v_3=0.2$ eV (blue points) and  $v_3=1$ eV (red points) for Normal Ordering (left) and Inverted Ordering (right).
 Lower: Scatter plots of the vev $v_3$ and the conversion probability $P$ in the  hybrid seesaw model with $m_{++}=800$ GeV (blue points) and  $m_{++}=3$ TeV (red points) for Normal Ordering (left) and Inverted Ordering (right).
 All the plotted points are allowed by the flavor and collider constraints.}%
\label{figure2}
\end{figure}

In this case, we assume all of the neutrino mass comes from the triplet Higgs, while the contribution from the right-handed neutrino can be negligible. In Fig. \ref{figure1} we show the scan result by requiring the neutrino mass satisfying the observation. Since there is an ambiguity in the neutrino mass ordering, we split our analysis in these two cases. The left panels show the preferred parameter space for neutrino mass Normal Ordering and the right panels that of Inverted Ordering. For the top two plots, we fixed the vacuum value of the triplet Higgs at 1 eV (red) and 0.2 eV (blue).  We find for $v_3= 1$ eV, the doubly-charged Higgs should be heavier than around 1.5 TeV to evade the limit from flavor constraints, particularly from $ \mu^+ \rightarrow e^- e^+ e^+$. In the case of $v_3= 0.2$ eV, the limit on the doubly-charged Higgs is stronger since a larger Yukawa coupling is needed to generate the neutrino mass for a smaller vacuum value of $\Delta$. We note that the preferred parameter space behaves as a band due to the uncertainty of the neutrino mass parameter. Finally, we find all the survived parameter space predicts a muonium-antimuonium conversion probability less than $10^{-15}$ which is beyond the reach of the future MACE experiment.

\subsection{Hybrid seesaw}
As shown in previous model, the strongest limit arises from the cLFV processes, such as $\mu^+ \to e^+ \gamma$ and $\mu^+ \to e^- e^+ e^+$. The reason is that from the PMNS matrix we can see the neutrino mixing is relatively large and a large off-diagonal terms in the neutrino mass matrix are needed. Since the neutrino mass matrix mainly originates from the triplet Higgs, it requires a large flavor changing coupling between the triplet Higgs and the leptons, inducing a large flavor changing effect. In the hybrid seesaw model, the flavor constrains could be weaker due to the contribution of off-diagonal terms from right-handed neutrino. Specifically, here we consider that all off-diagonal terms of the neutrino mass matrix originate from the right-handed neutrino, thus all of Yukawa couplings $y_\nu$ are fixed. Hence all the flavor constraints could be removed~\footnote{The contribution of the flavor changing observable from the right-handed neutrino can be safely removed assuming the right-handed neutrino is heavy enough. }. 

The final scanning results are shown in Fig.~\ref{fig:Pwithv2}. As before, the upper two scatter plots show the muonium-to-antimuonium conversion probability ${\cal P}(M\leftrightarrow \bar{M})$ as a function of the doubly-charged Higgs mass $m_{++}$ for both NO (left panel) and IO (right panel), while the lower two ones correspond to parameter spaces in ${\cal P}(M\leftrightarrow \bar{M})$ vs. $v_3$ plane. Compared with the previous Type-II seesaw case, it is clear that the allowed parameter spaces for the hybrid seesaw model are enlarged greatly, extending aggressively into the low doubly-charged Higgs mass region and the low triplet VEV $v_3$ region. Due to the relax of cLFV constraints in this hybrid model, the lower bounds on the doubly-charged Higgs mass for both NO and IO cases around 600~GeV are given by the collider searches for $\xi^{\pm\pm}$. As a result, it is seen that MACE can increase the detection region considerably. Concretely, for $v_3 = 0.2$~eV, 
MACE can detect doubly-charged Higgs as heavy as 3 TeV. On the other hand, for a fixed $m_{++}$, MACE has an ability to measure the parameter region with $v_3<1$~eV for both orderings, which is almost one order more sensitive than the PSI detector. The study in this subsection demonstrates the advantage and the necessity of MACE in the probing and constraining the triplet scalar in the seesaw models.      

\section{Summary and conclusion}
\label{sec:conclusion}

In the present paper, we have discussed prospects of the proposed MACE experiment to search for the muonium-to-antimuonium conversion process in the type-II and hybrid seesaw models, the latter of which is an extension of the type-II seesaw by including a single heavy right-handed neutrino. Note that the leading-order contributions to the $M-\bar{M}$ conversion probability in both models are induced at tree level by the doubly-charged scalar originated in the $SU(2)_L$ triplet scalar. Thus, MACE can effectively measure the parameter space related to this triplet scalar. By utilizing the high-quality slow muon sources at CSNS, together with other significant advances in the detector technologies, MACE has been shown to be able to improve the sensitivity to the conversion probability by almost three orders of magnitude compared with the existing PSI experiment accomplished more than 20 years ago. Therefore, a large uncharted model parameter space is expected to be probed by MACE, which is complementary to direct collider searches, neutrino oscillations and other cLFV channels such as $\mu^+ \to e^+ \gamma$ and $\mu^+ \to e^- e^+ e^+$. As a result, it is shown that MACE is not sensitive to the parameter space of interest surviving from the collider and flavor limits. On the other hand, for the hybrid seesaw model, MACE can reach unexplored parameter region, and thus provides useful information on the triplet scalar complementary to other cLFV experiments like MEG-II and Mu3e. In other words, if no positive signals are observed by MACE in the future, we shall be able to set a much more stringent upper bound on the mass of the doubly-charged scalar around 1.2 TeV for $v_3= 1$ eV in the case of Inverted Ordering, which will surpass the corresponding LHC limits even by taking into account the uncertainties involving the neutrino mass ordering or the VEV of the triplet scalar. Even for the NO, the prospect of MACE can reach a doubly-charged Higgs as heavy as 3 TeV for $v_3= 1$ eV. In summary, with the advent of MACE, the muonium-to-antimuonium conversion will become one of golden channels to study the cLFV physics, which, together with other flavor and collider searches, will shed light on the mystery of the neutrino masses.

\section*{Acknowledgement}
This work was supported in part by National Natural Science Foundation of China under Grant No. 12075326, and Guangdong Basic and Applied Basic Research Foundation under Grant No. 2019A1515012216, Original innovation project of Chinese Academy of Sciences (CAS) under Grant ZDBS-LY-SLH009 and CAS Center for Excellence in Particle Physics (CCEPP). JT appreciates fruitful discussions with the accelerator group of EMuS. CH is supported by SYSU startup funding. DH is supported by National Science Foundation of China (NSFC) under Grant No. 12005254. YZ is supported by National Science Foundation of China (NSFC) under Grant No. 11805001 and No. 11935001. We finally acknowledge Dr. Sampsa Vihonen's help in English proofreading.

\section*{Appendix}
\label{sec:appendix}
\subsection{Higgs Potential}
We consider the general Higgs potential for a doublet $\Phi = (h^+,h^0)^T$ and a triplet
Higgs $\Delta$~\cite{Chun:2003ej}.
\begin{eqnarray}
V(\Phi,\Delta)=&m^2(\Phi^\dagger\Phi)+\lambda_1(\Phi^\dagger\Phi)^2+M_\Delta^2\mathrm{Tr}(\Delta^\dagger\Delta)
+\lambda_2[\mathrm{Tr}(\Delta^\dagger\Delta)]^2
+\lambda_3\mathrm{Tr}(\Delta\Delta)\mathrm{Tr}(\Delta^\dagger\Delta^\dagger), \nonumber \\
&+\lambda_4(\Phi^\dagger\Phi)[\mathrm{Tr}(\Delta^\dagger\Delta)]+\lambda_5(\Phi^\dagger\Delta^\dagger\Delta\Phi)
+\lambda_6(\Phi^\dagger\Delta\Delta^\dagger\Phi)
+\frac{1}{2}\mu(\Phi^T\varepsilon\Delta^\dagger\Phi)
+\mbox{h.c.}
\end{eqnarray}
The breaking can be explicit when V($\Phi,\Delta$) is supposed to
contain the trilinear L-nonconserving term ($\Phi^T\Delta\Phi$). In
this case the masses of the new Higgs particles $M_\Delta$ in the
model can be arbitrary large. Collect the following soft lepton
number violating trilinear interaction between $\Delta$ and $\Phi$
from the Higgs potential:
\begin{eqnarray}
  {\cal L}_\mathrm{soft} & = & - \frac{1}{2}\mu \Phi^T \varepsilon
  \Delta^\dag \Phi + \mbox{h.c.}, \nonumber \\
  & = & -\frac{1}{2} \mu \left[ (h^+)^2 \xi^{--} - \sqrt{2} h^+ h^0 \xi^-
    - (h^0)^2 \xi^{0\dag} \right] + \mbox{h.c.}
\end{eqnarray}
where $\mu$ is an undetermined parameter. The minimum of Higgs
potential is presented in the following:
\begin{equation}
V_{\mathrm{min}}(\Phi,\Delta)=\frac{1}{2}
m^2v_1^2+M_\Delta^2v_3^2+\frac{1}{4}\lambda _1v_1^4+\lambda _2v_3^4
+\frac{1}{2}\lambda_4 v_1^2v_3^2+\frac{1}{2}\lambda _6v_1^2 v_3^2
+\frac{1}{2} \mu v_1^2v_3
\end{equation}
After the spontaneous symmetry breaking (SSB), solving the equation
$\frac{\partial V_{\mathrm{min}}(\Phi,\Delta)}{\partial v_i}=0$
leads to:
\begin{eqnarray}
  &\langle h^0 \rangle = \frac{v_1}{\sqrt{2}}, \quad
  \langle \xi^0 \rangle =v_3.\nonumber\\
  &v_1^2=-\frac{m^2}{\lambda _1}, \quad v_3=-\frac{\mu  v_1^2}{2 \left(2 M^2+v_1^2 \lambda _4+v_1^2 \lambda _6\right)}
  \label{eqn:v3}
\end{eqnarray}
where $\mu<0$, $m^2<0$ and $M_\Delta^2>0$. When $|\mu|$ is small
enough compared with the weak scale, the smallness of neutrino
masses is attributed to the tiny $\mu$, which is estimated at the eV
scale in the case where $y_N \sim O(1)$ and $M_\Delta \sim v_1$. For
completeness, the charged Higgs masses are listed below:
\begin{eqnarray}
 m_{+}&=&M_\Delta^2+\frac{1}{2}(\lambda_4+\frac{\lambda_5+\lambda_6}{2})v_1^2\\
 m_{++}&=&M_\Delta^2+\frac{1}{2}(\lambda_4+\lambda_5)v_1^2
\end{eqnarray}

\subsection{Gauge Sector}
Now let us move to gauge couplings $D_\mu\xi D^\mu\xi$.
Since there is a triplet scalar, we have to use 3-dimension
representation of SU(2) group.
\begin{equation}
  T_1=\left(
\begin{array}{lll}
 0 & \frac{1}{\sqrt{2}} & 0 \\
 \frac{1}{\sqrt{2}} & 0 & \frac{1}{\sqrt{2}} \\
 0 & \frac{1}{\sqrt{2}} & 0
\end{array}
\right),\quad T_2=\left(
\begin{array}{lll}
 0 & -\frac{i}{\sqrt{2}} & 0 \\
 \frac{i}{\sqrt{2}} & 0 & -\frac{i}{\sqrt{2}} \\
 0 & \frac{i}{\sqrt{2}} & 0
\end{array}
\right), \quad T_3=\left(
\begin{array}{ccc}
 1 & 0 & 0 \\
 0 & 0 & 0 \\
 0 & 0 & -1
\end{array}
\right)
\end{equation}
\begin{equation}
  T^+=T^1+iT^2=\left(
\begin{array}{ccc}
 0 & \sqrt{2} & 0 \\
 0 & 0 & \sqrt{2} \\
 0 & 0 & 0
\end{array}
\right),\quad T^-=T^1-iT^2=\left(
\begin{array}{ccc}
 0 & 0 & 0 \\
 \sqrt{2} & 0 & 0 \\
 0 & \sqrt{2} & 0
\end{array}
\right)
\end{equation}
\begin{equation}
  D_\mu\xi=\partial_\mu-ig_1Y_\xi B_\mu-ig_2A_\mu^aT^a=\left(
\begin{array}{l}
  \left(\partial _{\mu }-i B_{\mu } g_1-i
   A^3_{\mu } g_2\right)\xi ^{\text{++}}-ig_2 W^+_{\mu }\xi ^+ \\
  \left(\partial _{\mu }-i B_{\mu } g_1\right)\xi ^+-ig_2
  W^-_{\mu }\xi ^{\text{++}}-ig_2 W^+_{\mu }\xi ^0 \\
  \left(\partial _{\mu }-i B_{\mu } g_1+i A^3_{\mu }
   g_2\right)\xi ^0-ig_2 W^-_{\mu }\xi ^+
\end{array}
\right)
\end{equation}

\begin{equation}
\begin{aligned}
W_{\mu}^\pm&=\frac{1}{\sqrt{2}}(A_{\mu}^1\mp iA_{\mu}^2)\\
s_w&\equiv\sin\theta_w=\frac{g_1}{\sqrt{g_1^2+g_2^2}}\\
c_w&\equiv\cos\theta_w=\frac{g_2}{\sqrt{g_1^2+g_2^2}}\\
A_{\mu}^3&=Z_{\mu}^0\cos\theta_w+A_{\mu}\sin\theta_w\\
B_{\mu}&=-Z_{\mu}^0\sin\theta_w+A_{\mu}\cos\theta_w
\end{aligned}
\end{equation}
After spontaneous symmetry breaking, $\Delta$ develops a vacuum expectation value. If physical vector bosons are introduced, we have
\begin{equation}
  D^{\mu*}(\xi+\langle\xi\rangle)=\left(
\begin{array}{l}
 i W^{-\mu} \xi ^- g_2+ \left[i g_2
   \left(Z^{\text{0$\mu $}} c_w+A^{\mu } s_w\right)+i g_1 \left(A^{\mu }
   c_w-Z^{\text{0$\mu $}} s_w\right)+\partial ^{\mu }\right]\xi ^{--} \\
 i A^{\mu } \xi ^- c_w g_1+ig_2 \left(W^{\text{+$\mu $}} \xi
   ^{--}+W^{-\mu } \xi ^{\text{0*}}\right) +
   \left(\partial ^{\mu }-i Z^{\text{0$\mu $}} g_1 s_w\right)\xi ^- \\
 i W^{\text{+$\mu $}} \xi ^- g_2+ \left[-i g_2
   \left(Z^{\text{0$\mu $}} c_w+A^{\mu } s_w\right)+i g_1 \left(A^{\mu }
   c_w-Z^{\text{0$\mu $}} s_w\right)+\partial ^{\mu }\right]\xi ^{\text{0*}}
\end{array}
\right)
\end{equation}
They will contribute to gauge boson masses,
\begin{equation}
\begin{split}
  |D_\mu\langle\xi\rangle|^2&=W_{\mu }^+ W^{-\mu} g_2^2 v_3^2+Z_{\mu }^0 Z^{\text{0$\mu $}}
   \left(c_w^2 g_2^2+2 c_w g_1 s_w g_2+g_1^2 s_w^2\right) v_3^2\\
   &\quad+Z_{\mu }^0
   A^{\mu } \left(-g_1 g_2 c_w^2-g_1^2 s_w c_w+g_2^2 s_w c_w+g_1 g_2
   s_w^2-g_1 g_2 c_w^2-g_1^2 s_w
   c_w+g_2^2 s_w c_w+g_1 g_2 s_w^2\right) v_3^2\\
   &\quad+A_{\mu } A^{\mu } \left(c_w^2
   g_1^2-2 c_w g_2 s_w g_1+g_2^2 s_w^2\right) v_3^2\\
   &=g_2^2v_3^2W^{+\mu}W_\mu^{-}+\frac{g_2^2v_3^2}{\cos^2\theta_w}Z^{0\mu}Z_\mu^{0}
\end{split}
\end{equation}
In addition, the traditional SM contains gauge mass terms:
\begin{equation}
\mathcal{L}_\mathrm{GM}=\frac{1}{4}g_2^2v_1^2W_{\mu}^{+}W^{-\mu}
+\frac{1}{8}\frac{g_2^2v_1^2}{\cos^2\theta_w}Z_{\mu}^0Z^{0\mu}
\end{equation}
which violates the custodial symmetry such as
\begin{equation}
  \rho=\frac{M_W^2}{M_Z^2\cos^2\theta_w}=\frac{\frac{1}{4}g_2^2v_1^2+g_2^2v_3^2}
  {\frac{1}{4}g_2^2v_1^2+2g_2^2v_3}=1-\frac{v_3^2}{\frac{1}{4}v_1^2+2v_3^2}
  \label{eqn:rho}
\end{equation}
As $|\mu|$ is small enough so that $v_3\ll v_1$, $\rho\approx 1$
meets the requirement from electroweak precision experiments whereas
detectable lepton flavor violating processes are expected. Since the
lepton number is restored for $\mu=0$, it may be natural to have a
small $\mu$ as a consequence of a tiny lepton number violation. In
fact, there is a tree-level lepton flavor violating process
$e^-\mu^+\longrightarrow e^+\mu^-$ mediated by $\xi^{++}$ due to the
interaction term $\bar{e}^c_i P_Le_j \xi^{++}$~\cite{Chang:1989uk}
just as the related term of Eq.~(\ref{Yuk}).

\bibliographystyle{unsrt}	
\bibliography{main}	

\end{document}